\documentclass[amsmath,pra,reprint,showpacs,superscriptaddress,aps,10pt]{revtex4-1}
\usepackage{amsmath,amssymb,graphicx}

\usepackage{color}
\usepackage{ulem}

\begin{document}

\title{Back-Scattering Properties of a Waveguide-Coupled Array of Atoms in the Strongly Non-Paraxial Regime}
\author{D. Reitz}
\author{C. Sayrin}
\author{B. Albrecht}
\affiliation{Vienna Center for Quantum Science and Technology, Atominstitut, TU Wien, Stadionallee 2, 1020 Vienna, Austria}
\author{I. Mazets}
\affiliation{Vienna Center for Quantum Science and Technology, Atominstitut, TU Wien, Stadionallee 2, 1020 Vienna, Austria}
\affiliation{Ioffe Physico-Technical Institute of the Russian Academy of Sciences, 194021 St. Petersburg, Russia}
\author{R. Mitsch}
\author{P. Schneeweiss}
\author{A. Rauschenbeutel}\email{Corresponding author: arno.rauschenbeutel@ati.ac.at}
\affiliation{Vienna Center for Quantum Science and Technology, Atominstitut, TU Wien, Stadionallee 2, 1020 Vienna, Austria}

\begin{abstract}
We experimentally investigate the back-scattering properties of an array of atoms that is evanescently coupled to an optical nanofiber in the strongly non-paraxial regime. We observe that the power and the polarization of the back-scattered light depend on the nanofiber-guided excitation field in a way that significantly deviates from the predictions of a simple model based on two-level atoms and a scalar waveguide. Even though it has been widely used in previous experimental and theoretical studies of waveguide-coupled quantum emitters, this simple model is thus in general not adequate even for a qualitative description of such systems. We develop an ab initio model which includes the multi-level structure of the atoms and the full vectorial properties of the guided field and find very good agreement with our data.
\end{abstract}

\pacs{42.50.Ct, 42.81.Qb, 37.10.Jk}

\maketitle

Recently, there has been growing theoretical and experimental interest in the physics of quantum emitters coupled to optical waveguides. Various phenomena have been predicted, including self-organization of atoms~\cite{Chang13,Griesser13}, cavity quantum electrodynamics with atomic mirrors~\cite{Chang12}, and the formation of a Tonks-Girardeau gas of photons~\cite{Chang08,Kiffner10}. Atom-mediated directional emission~\cite{LeKien08,Zoubi10c} and quantum transport of strongly interacting photons~\cite{Hafezi12b} have been theoretically studied, and the non-radiative interaction and entanglement between distant atoms along the waveguide has been proposed~\cite{Shahmoon13b}. Most of these theoretical works disregard the vectorial character of the waveguide modes or approximate the emitters as two-level systems. However, these approximations are not necessarily justified and the predicted phenomena may not prevail in a real world scenario.

We study the scattering properties of an ensemble of laser-cooled cesium atoms trapped in two linear arrays in the evanescent field around an optical nanofiber that  realizes a single-mode waveguide~\cite{LeKien04,Vetsch10, Goban12}. We find qualitative deviations from the predictions of the simplified model of two-level atoms coupled to a scalar radiation field. In particular, a quantitative description has to consider the reduction of the overlap between counter-propagating waveguide modes due to their non-paraxial character~\cite{Junge13} which leads to a counter-intuitive back-scattering signal. Moreover, the multi-level structure of the atoms leads to inelastic scattering, thereby coupling modes which are orthogonal in the full vectorial description. We expect our findings to improve the understanding of atom--waveguide systems and of other quantum optics experiments in the non-paraxial regime, like atoms coupled to plasmonic structures~\cite{Stehle11}, nanophotonic cavities~\cite{Thompson13b} or optical microtraps~\cite{Kaufman12,Thompson13}.

\begin{figure}
\centerline{\includegraphics[width=0.95\columnwidth]{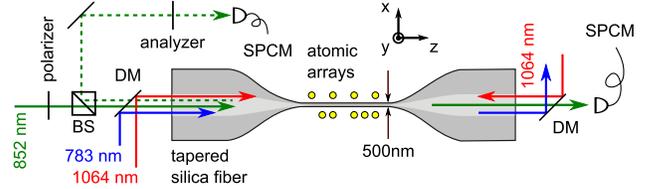}}
\caption{Sketch of the experimental set-up including the tapered optical fiber and the trapping (blue and red lines) and probe (green line) laser fields. A polarizer sets the linear polarization of the probe field. The probe transmission is recorded using a single photon counting module (SPCM). A fraction of the back-scattered light is separated from the forward propagating fields by a non-polarizing beam splitter (BS), passes through an analyzer, and is sent onto another SPCM. Dichroic mirrors (DM) and interference filters (not shown) prevent the trapping fields from reaching the SPCMs.}
\label{fig:setup}
\end{figure}

The experimental set-up is depicted in Fig.~\ref{fig:setup}. Laser-cooled Cs atoms are confined in the 3D Lamb-Dicke regime using a nanofiber-based two-color dipole trap~\cite{Vetsch12}. The optical nanofiber has a nominal radius $a=250~$nm and is realized as the waist of a tapered optical fiber~\cite{Brambilla10} which enables close to unity coupling efficiency between the standard fiber and the nanofiber waist. For all optical wavelengths involved in this experiment, the optical nanofiber is sufficiently thin to only guide the fundamental hybrid HE$_{11}$ mode~\cite{LeKien04b}. Trapping of the atoms is achieved by using a red-detuned standing-wave with a free-space wavelength of $1064\,{\rm nm}$ and a power of $2\times1.4\,{\rm mW}$ combined with a blue-detuned traveling-wave with a wavelength of $783\,{\rm nm}$ and a power of $13.3\,{\rm mW}$. The atoms are located $200\,{\rm nm}$ above the nanofiber surface in two diametric linear arrays of potential wells. The trap frequencies in all three (radial, axial, and azimuthal) directions are about $100\,\mathrm{kHz}$. A few hundred atoms are typically loaded into the trap, with at most one atom per trapping site~\cite{Vetsch12}.

In order to probe the atoms, a linearly polarized light field, resonant with the AC-Stark shifted $F=4 \to F'=5$ transition of the Cs D2 line (free-space wavelength $\lambda=852\,{\rm nm}$), is launched into the fiber in the forward $(+z)$ direction. A polarizer determines its polarization. The transmission of the probe field is measured with a single photon counting module (SPCM). A fraction of the back-scattered light is separated from the fields propagating in the forward direction by a non-polarizing beam splitter, passes through a polarization analyzer, and is detected with another SPCM.  Two Berek compensators, one in front of the fiber and one in front of the analyzer, compensate for the parasitic birefringence of the fiber~\cite{Vetsch12}. In this way, linearly polarized free-space modes are mapped on quasi-linearly polarized HE$_{11}$ modes in the nanofiber~\cite{LeKien04b} and vice versa. The polarizer and analyzer thus allow one to selectively prepare and measure any quasi-linearly polarized nanofiber mode, respectively.

The large refractive index contrast between the silica optical nanofiber and the surrounding vacuum transversally confines the nanofiber modes to less than $\lambda^2$, thereby making them strongly non-paraxial~\cite{LeKien04b}: Their evanescent field locally exhibits a significant longitudinal polarization component which is $\pi/2$-phase shifted with respect to the transversal components. The total intensity as well as the longitudinal field component and thus the polarization vary azimuthally.

\begin{figure}
\centerline{\includegraphics[width=0.95\columnwidth]{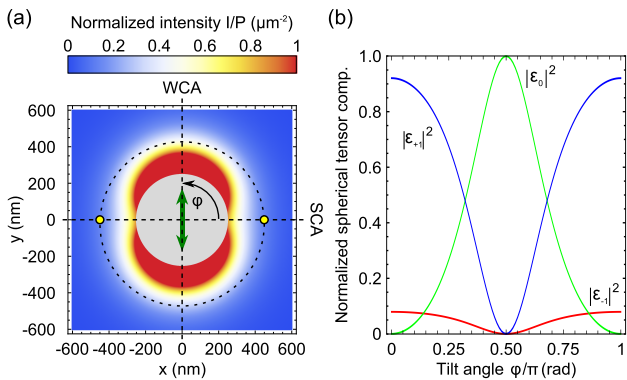}}
\caption{
Characteristics of a quasi-linearly polarized HE$_{11}$ mode propagating in the $(+z)$-direction, calculated for $\lambda=852\,{\rm nm}$. (a) Density plot of the intensity profile. The trapped atoms are indicated by the yellow dots and the WCA and SCA by two dashed lines. Here, the main polarization axis (green double arrow) coincides with the WCA ($\varphi=\pi/2$). (b) Modulus squared of the normalized spherical tensor components $(\mathcal{E}_{-1},\;\mathcal{E}_{0},\;\mathcal{E}_{+1})/|\boldsymbol{\mathcal{E}}|$ of the field for the right-hand-side atom as a function of $\varphi$, plotted in red, green, and blue, respectively. The components $\mathcal{E}_{-1}$ and $\mathcal{E}_{+1}$ have to be interchanged for the left-hand-side atom.
}\label{fig:modes}
\end{figure}

The intensity profile of a quasi-linearly polarized nanofiber-guided field is shown in Fig.~\ref{fig:modes}(a). Its main transversal polarization component and the plane containing the atoms enclose an angle $\varphi$. In Fig.~\ref{fig:modes}(a), this main polarization axis is aligned along the $y$-direction. The azimuthal minima of the intensity then coincide with the position of the trapped atoms, and the coupling between the atoms and the field is minimal. Thus, we label the $y$-axis as the weak coupling axis (WCA). Accordingly, the $x$-axis is called the strong coupling axis (SCA), where the intensity is 2.8 times larger than on the WCA. In Fig.~\ref{fig:modes}(b), we plot the modulus square of the normalized spherical tensor components $(\mathcal{E}_{-1},\;\mathcal{E}_{0},\;\mathcal{E}_{+1})/|\boldsymbol{\mathcal{E}}|$~\cite{Shore1990} of the probe field $\boldsymbol{\mathcal{E}}$ at the position of the atoms as function of $\varphi$. We take the $y$-axis as the quantization axis: $\mathcal{E}_{0}=\mathcal{E}_{y}$, $\mathcal{E}_{\pm 1} = \pm(\mathcal{E}_{x} \pm i \mathcal{E}_{z})/\sqrt{2}$. If the polarization is aligned along the WCA ($\varphi=\pi/2$), the field is purely linear at the position of the atoms and drives $\pi$-transitions. If $\varphi\in\{0,\pi\}$, the polarization is almost circular ($|\mathcal{E}_{\pm 1}/\mathcal{E}|=0.96$), and $\mathcal{E}_{0}=0$. The field then essentially drives $\sigma^-$ ($\sigma^+$) transitions for the atoms located at $x<0$ ($x>0$). If the probe field propagates in the backward $(-z)$ direction, the situation is reversed~\cite{Junge13} and $\sigma^-$ ($\sigma^+$) transitions are driven for atoms located at $x>0$ ($x<0$). This results in a reduced overlap of about $0.29$ between the forward and backward propagating SCA modes, whereas there is full overlap for the WCA modes~\cite{Junge13}.

We record the transmission and back-scattering of a 10-$\mu$s nanofiber-guided probe pulse, with a power $P^+_{i}$, that is quasi-linearly polarized along either the WCA ($i={\rm w}$) or the SCA ($i={\rm s}$). The transmitted and back-scattered powers are measured in parallel with the two SPCMs. The photon counts are recorded with a 100-ns binning time. Taking into account the experimental imperfections, we convert the count rates to optical powers in the nanofiber. We observe that the back-scattered power remains constant during the first $200\,{\rm ns}$ of the pulse: Neither the motion of the atoms in the trap nor optical pumping play a significant role within this time interval. Therefore, we average the back-scattering signal over the first two time-bins only. The transmission remains constant over the first 500~ns and is thus averaged over the first five time-bins.

\begin{figure}
\centerline{\includegraphics[width=0.95\columnwidth]{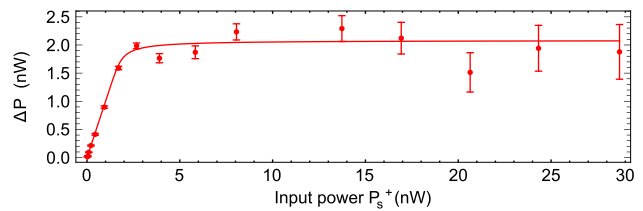}}
\caption{Difference $\Delta P$ between the transmitted powers recorded in the absence and presence of the trapped atoms as a function of $P^+_{\rm s}$. Each data point is the average of 80 experimental runs. The probe light is polarized along the SCA. The red solid line is a fit, see text.}
\label{fig:transmission}
\end{figure}

Figure~\ref{fig:transmission} shows an outcome of a transmission measurement. We plot the difference $\Delta P$ between the transmitted powers recorded in the absence and presence of the trapped atoms. A clear saturation is visible. Following \cite{Vetsch10}, we determine the number of trapped atoms $N$ by fitting the data using a generalized Beer-Lambert law~\cite{Stenholm1984} which describes the propagation of a light field through an absorptive and saturable medium. It is given by $N=\Delta P_\infty/P_{\rm Cs}$, where $\Delta P_\infty$ is the asymptotic value of $\Delta P$. The maximum scattered power per Cs atom $P_{\rm Cs}=\hbar \omega \Gamma /2 = 3.8\,{\rm pW}$ is independent of the polarization of the probe field. Here,  $\omega$ is the angular frequency of the optical transition and $\Gamma=2\pi\times5.2$~MHz is the excited state decay rate.

\begin{figure}
\includegraphics[width=0.95\columnwidth]{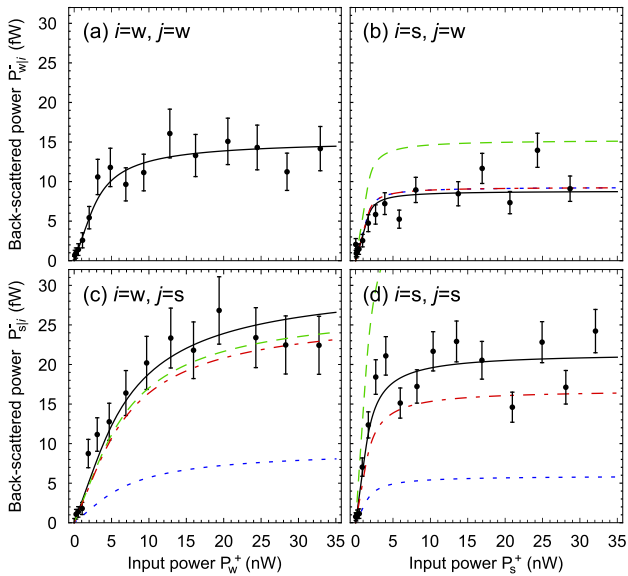}
\caption{Measured back-scattered power divided by the number of trapped atoms as a function of $P^+_{i}$ for the four polarizer/analyzer settings. Each data point is the average of 80 experimental runs. The solid black lines are fits obtained by solving Eq.~\eqref{eq:DiffBack}. The dashed-dotted red, dashed green and dotted blue lines are calculated using the ab initio, the ``intensity-only'' and the ``polarization-only'' models, respectively (see text).}
\label{fig:data}
\end{figure}

In Fig.~\ref{fig:data}, we plot the back-scattered power $P^-_{j|i}$ as function of $P^+_{i}$. It is measured for the four possible settings of the polarizer ($i\in$\{s,w\}) and the analyzer ($j\in$\{s,w\}). The power $P^-_{j|i}$ is normalized to the number of atoms $N$, measured within the same experimental realization. This normalization is motivated by the assumption that the back-scattered power is proportional to $N$ in the regime of full saturation of the atomic sample~\cite{supplemental}. In all measurements, the total back-scattered power $N P^-_{j|i}$ is three orders of magnitude smaller than~$P^+_{i}$. For any choice of \{\textit{j,i}\}, $P^-_{j|i}$ shows a clear saturation. For an input polarization along the SCA, saturation occurs at lower input powers than for an input polarization along the WCA. This is consistent with the intensity map shown in Fig.~\ref{fig:modes}(a). 

The significant back-scattered powers for the crossed polarizer--analyzer setting [Figs.~\ref{fig:data}(b) and (c)] reveal the presence of an inelastic scattering mechanism that changes both the polarization of the scattered light and the internal state of the atom. It can only be understood when considering the hyperfine and Zeeman sub-structure of the atom. Remarkably, all measurements in Fig.~\ref{fig:data} level off at different values of the back-scattered power per atom, again demonstrating scattering dynamics beyond what is expected for a two-level atom coupled to a scalar waveguide. More precisely, for both input polarizations, the asymptotic back-scattered powers are larger with the analyzer along the SCA than along the WCA. The larger intensity of the SCA mode at the position of the atoms and the correspondingly larger coupling strength partly explain this difference. 
Furthermore, we observe different asymptotic power levels for configurations with the same output but different input modes [Figs.~\ref{fig:data}(a) and (b) as well as Figs.~\ref{fig:data}(c) and (d)]. Given that, far above saturation, emission rates do not depend on the probe intensity anymore, we conclude that polarization effects must be at the origin of this difference. Indeed, there is a finite overlap between the polarization of the field emitted by the atoms and the fiber modes. The polarization of the emitted field depends on the polarization of the probe that excites the atoms, and so does this overlap. This also explains the surprising fact that the highest count rates are measured with the probe aligned along the WCA where its intensity at the position of the atoms is lowest. 

In order to quantitatively understand our experimental observations, we develop a model for the $z$-dependent power of the back-scattered nanofiber-guided field, $\mathcal{P}^-_{j|i}(z)$. At every position $z$ along the nanofiber, the optical power of the probe, $\mathcal{P}^+_{i}(z)$, is given by a generalized Beer-Lambert law, accounting for the saturable absorption by the atoms. Given that $\mathcal{P}^-_{j|i}(z)\ll\mathcal{P}^+_{i}(z)$, we assume that the saturation level of the atoms is solely determined by $\mathcal{P}^+_{i}(z)$~\cite{supplemental}. The backward propagating power then obeys
\begin{equation}
\label{eq:DiffBack}
\frac{d}{d z} \mathcal{P}^-_{j|i}(z) = \frac{n \sigma_j/A^{\rm eff}_j}{1+s_{i}(z)}  \mathcal{P}^-_{j|i}(z) 
 - \frac{s_{i}(z)}{1+s_{i}(z)}\, n\, P_{j|i}^{-,\rm max}~,
\end{equation}
where $n\propto N$ is the atomic line density, $\sigma_j$ is the atomic cross section for the interaction with the back-scattered nanofiber-guided field,  and $A^{\rm eff}_j$ denotes the effective mode area \cite{LeKien06c}, see Fig.~\ref{fig:modes}(a). 
The first term on the r.h.s.~of Eq.~\eqref{eq:DiffBack} describes the damping of the backward propagating light due to absorption by the atoms. The position-dependent saturation parameter is given by $s_{i}(z)= \mathcal{P}^+_{i}(z)/P^{\rm sat}_{i}$. Here, $P^{\rm sat}_{i}$ is the optical power of the nanofiber-guided field that is required to reach saturation intensity at the position of the atoms~\cite{supplemental}.
The interaction cross section is then given by $\sigma_j=A^{\rm eff}_j (P_{\rm Cs}/P^{\rm sat}_{j})$. The second term on the r.h.s.~of Eq.~(\ref{eq:DiffBack}) accounts for the saturable emission into the backward mode. It is proportional to $P_{j|i}^{-,\rm max}$, defined as the back-scattered power per atom at full saturation.

Using the analytical solution for the back-scattered power $P^-_{j|i} = \mathcal{P}^-_{j|i}(z=0)$ as a function of $P^+_{i}$, we model the data from Fig.~\ref{fig:data} using $P_{j|i}^{-, \rm max}$ and $N$ as the only free parameters. We fit the back-scattering data sets simultaneously with the corresponding transmission signals~\cite{supplemental} for each given polarizer and analyzer setting. The results  shown as solid red and black lines in Fig.~\ref{fig:transmission} and \ref{fig:data}, respectively, are in very good agreement with the data. The fitted values of $P_{j|i}^{-,\rm max}$ and $N$ are given in Tab.~\ref{tab:backscatteringrates}.
\begin{table}
\begin{tabular}{c|c|c|c}
Setting & $N$ & $P_{j|i}^{-,\rm max}$ (fW) & $P_{j|i}^{-,\rm theo}$ (fW) \\
\hline \hline
$i=\textrm{w}, j=\textrm{w}$ & $739\pm 42$ & $15.3 \pm 0.9$ & $20$\\
$i=\textrm{s}, j=\textrm{w}$ & $546\pm 38$ & $8.8 \pm 0.7$ & $12$\\
$i=\textrm{w}, j=\textrm{s}$ & $648\pm 38$ & $30.7 \pm 1.5$ & $35$\\
$i=\textrm{s}, j=\textrm{s}$ & $408\pm 32$ & $21.6 \pm 1.6$ & $22$\\
\end{tabular}

\caption{Fitted values of $P_{j|i}^{-,\rm max}$ and $N$, and results of the ab initio calculation $P_{j|i}^{-,\rm theo}$ (see text).}
\label{tab:backscatteringrates}
\end{table}

We now compare the fit results for the maximum back-scattered power per atom with the results of an ab initio calculation. For this, we take the local intensity and local polarizations of the fiber-guided modes at the position of the trapped atoms into account. We assume an initial statistical mixture of all $F=4$ Zeeman ground states with equal populations and calculate the density matrix $\rho_{i}$ of the atom after absorption of a photon from mode $i$ that is resonant with the $F=4 \to F^\prime = 5$ transition. We consider strong saturation for which the Zeeman state-dependent level shifts induced by the trapping light fields~\cite{LeKien13c} can be neglected. Using the formalism outlined in \cite{LeKien05c}, we then calculate the scattered power $P_{j|i}^{-,\rm cond}$ into the backward-propagating mode $j$, conditioned on this absorption. It is given by
\begin{align}
\label{eq:EmRate}
P_{j|i}^{-,\rm cond} &= \hbar \omega \,\mathrm{Tr}\left(\Gamma_{j}\rho_{i}\right) .
\end{align}
Here, the non-diagonal matrix $\Gamma_{j}$ describes the spontaneous emission into the guided mode $j$~\footnote{Note that our definition of $\Gamma$ is transposed with respect to that of Ref.~\cite{LeKien05c}}.
The value $P_{j|i}^{-,\rm cond}$ is calculated for a fully excited atom. Considering that the steady-state excited state population cannot be higher than $0.5$, one expects a fitted power   $P_{j|i}^{-,\rm max}\lesssim P_{j|i}^{-,\rm theo}= 0.5 P_{j|i}^{-,\rm cond}$. The values of $P_{j|i}^{-,\rm theo}$ are summarized in Tab.~\ref{tab:backscatteringrates}. Remarkably, they are in good agreement with the measured values.

We now show that the data can indeed only be satisfactorily explained by a model that takes both the intensity and polarization maps of the nanofiber modes into account. For this purpose, we establish two simpler models: an ``intensity-only'' model where the local polarization of the nanofiber modes is neglected, obtained by replacing $\rho_{i}$ in Eq.~\eqref{eq:EmRate} by an equiprobable statistical mixture of all sub-Zeeman states of the $F'=5$ manifold, and a ``polarization-only'' model where the intensity profile of the fiber modes is neglected, obtained by replacing the matrix $\Gamma_{j}$ in Eq.~\eqref{eq:EmRate} by the one calculated for an atom in free-space. These models then allow us to predict the ratios of the asymptotic back-scattered powers. Taking the fitted value for $P_{\rm w|w}^{-,\rm max}$ as a reference, see Fig.~\ref{fig:data}(a), we calculate the values for the three other configurations according to $P_{j|i}^{-,\rm max} = \left[P_{j|i}^{-,\rm theo}/P_{\rm w|w}^{-,\rm theo}\right] P_{\rm w|w}^{-,\rm max}$.
The power $P^-_{j|i}$ is then calculated by solving Eq.~\eqref{eq:DiffBack} with this value of $P_{j|i}^{-,\rm max}$ and the fitted number of atoms $N$. 

From Fig.~\ref{fig:data}(b), it is conspicuous that the ``polarization-only'' model is well-suited to predict how $P_{j|i}^{-,\rm max}$ is modified when rotating the input polarization while leaving the analyzer unchanged. This model predicts that the polarization of the emitted photon is different for the $i=\rm s$ and the $i=\rm w$ configurations. Its overlap with the polarization of the backward propagating $j=\rm w$ mode is thus modified and so is $P_{{\rm w}|i}^{-,\rm max}$. From Fig.~\ref{fig:data}(c) however, it is apparent that the ``intensity-only'' model is more accurate than the ``polarization-only'' model when it comes to comparing the values of $P_{j|i}^{-,\rm max}$ that correspond to two different output modes, $j=\rm w$ and $j=\rm s$, while leaving the input polarization unchanged. In this situation, the modification of the effective mode area explains most of the modification of the back-scattered power. Both simple models however fail to match the data well for the \{s,s\} configuration [Fig.~\ref{fig:data}(d)], i.e., when both the input and output polarizations are changed with respect to the \{w,w\} reference settings. In this case, both the intensity profile and the polarization of the nanofiber modes need to be considered. The predictions of the full model, again referenced to $P_{\rm w|w}^{-,\rm max}$, exhibit good agreement with the data for all three other configurations.

Summarizing, we studied the back-scattering of cold Cs atoms trapped in two diametric linear arrays that are coupled to an optical nanofiber. We found that both the polarization and the intensity map of the nanofiber-guided modes as well as the multilevel structure of the atoms have to be taken into account in order to reach a deeper understanding of the scattering properties of the system. Neglecting these effects, as it has been commonly done both in theoretical investigations and experimental analyses of emitter--waveguide-systems so far, may lead to quantitatively wrong predictions and even qualitative discrepancies between the theory and the experimental observations. For instance, we made the counter-intuitive observation that the back-scattered power can be significantly larger when choosing the input polarization that minimizes the intensity at the position of the atoms to a third of its peak value, cf.~left vs.~right columns in Fig.~\ref{fig:data}. 

In the experimental situation realized here, collective effects like sub- and superradiance can be neglected and the scattering properties are those of an ensemble of independent scatterers. However, the nature of the observed effects leads us to conclude that they will also modify the collective scattering properties of denser ensembles or of ensembles that fulfill the Bragg condition. Finally, given that the longitudinal polarization component of the light plays a decisive role in the modification of the scattering properties observed here, similar phenomena should occur in other cases of strongly non-paraxial light--matter coupling like strongly focused light fields~\cite{Kaufman12,Thompson13}, plasmonics~\cite{Stehle11}, or nanophotonic systems~\cite{Thompson13b}.

We acknowledge financial support by the Austrian Science Fund (FWF, SFB NextLite project No. F~4908-N23, DK CoQuS project No. W~1210-N16, and project No. P~25329-N17). C.~S.~acknowledges support by the EC (Marie Curie IEF Grant 328545). D.~R.~and C.~S.~contributed equally to this work.

\bibliography{Reflection}

\end{document}


\title{Supplementary Information: Back-Scattering Properties of a Waveguide-Coupled Array of Atoms in the Strongly Non-Paraxial Regime}
\author{D. Reitz}
\author{C. Sayrin}
\author{B. Albrecht}
\affiliation{Vienna Center for Quantum Science and Technology, Atominstitut, TU Wien, Stadionallee 2, 1020 Vienna, Austria}
\author{I. Mazets}
\affiliation{Vienna Center for Quantum Science and Technology, Atominstitut, TU Wien, Stadionallee 2, 1020 Vienna, Austria}
\affiliation{Ioffe Physico-Technical Institute of the Russian Academy of Sciences, 194021 St. Petersburg, Russia}
\author{R. Mitsch}
\author{P. Schneeweiss}
\author{A. Rauschenbeutel}
\affiliation{Vienna Center for Quantum Science and Technology, Atominstitut, TU Wien, Stadionallee 2, 1020 Vienna, Austria}

\date{\today}
\maketitle

\section*{Collective Scattering}
The findings presented in the main manuscript have been obtained with a nanofiber-based two-color dipole trap similar to the one in Refs.~\cite{Vetsch10,Vetsch12}: Two diametric, linear arrays of trapping sites are formed 200~nm above the nanofiber surface. The atoms are transferred into the nanofiber-based trap from a magneto-optical trap using an intermediate stage of optical molasses cooling. For our experimental parameters, the so-called collisional blockade effect limits the number of atoms per site to one~\cite{Schlosser02}. Thus, the average filling factor is $\lesssim 0.5$~\cite{Vetsch12} and the trapped atoms are randomly distributed over the linear arrays of trapping sites. In combination with the fact that the distance between neighboring trapping sites is incommensurate with the probe wavelength in the fiber mode, the fields that are elastically back-scattered by the individual atoms effectively have random phases and therefore add up incoherently. In the case of full saturation of the atomic transition, the scattering is inelastic~\cite{Tannoudji10} and the individual fields add up incoherently as well. Thus, we do not expect any collective effects in the back-scattering of the atomic ensemble. 

Moreover, we do not expect a significant modification of the lifetime of the excited states due to sub- or superradiance with respect to the free-space modes: The smallest distance between any two atoms is larger than half the probe wavelength, meaning that there is conclusive which-way-information for scattering into free space. 

Finally, the modification of the excited state decay rate of an individual atom due to its close proximity of 200~nm to the nanofiber surface is $\lesssim 5$~\% for our experimental parameters~\cite{LeKien06c} and therefore neglected in our analysis.

\section*{Saturation of the atoms and absorption of the backward propagating light}
In Eq.~(1) of the main text, we assumed that the saturation level of the atoms is solely determined by the forward propagating light field $\mathcal{P}^+_{i}(z)$. In a two-level system, this assumption is obviously justified: Both the forward and the backward propagating light fields couple to the only available transition between the ground and the excited state. Moreover, the backward propagating power is much smaller than the forward propagating power, $\mathcal{P}^-_{j|i}(z)\ll \mathcal{P}^+_{i}(z)$. Thus the backward propagating light couples to a saturated transition but does not significantly contribute to its saturation. In the following, we will show that this assumption remains valid to a good approximation even in the case of transitions between the $F=4$ ground state manifold and the $F'=5$ excited state manifold  of the Cs atoms used in the experiment.

If the polarization of the forward propagating light field is set to the SCA, the fiber-guided mode at the position of the atoms is elliptically polarized, i.e., it has both $\sigma^+$ and $\sigma^-$ components. It thus couples the $F=4$ manifold to all Zeeman sub-states of the $F'=5$ manifold, and the situation becomes similar to the case of the two-level system once all transitions are saturated.

\begin{figure}
\centerline{\includegraphics[width=0.95\columnwidth]{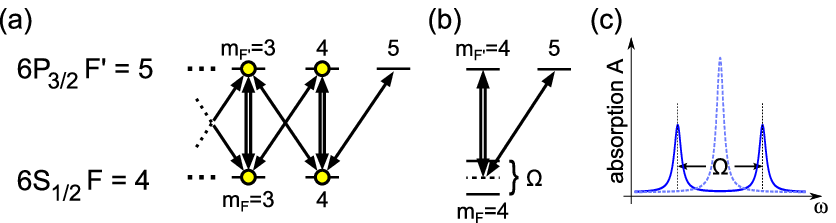}}
\caption{(a) Level structure and transitions driven by the forward propagating field aligned along the WCA (double line arrows) and the backward propagating field aligned along the SCA (single line arrows). The atomic population in the regime of strong saturation, $s_{\rm w}\gg1$, is indicated by yellow dots. (b) The strong forward propagating field induces an Autler-Townes splitting of the $m_F=4$ ground state. This results in a reduction of the absorption $A$ of the backward propagating field on the $m_F=4\to m_{F'}=5$ transition. (c)~Schematic spectrum of the absorption $A(\omega)$  of a weak field at angular frequency $\omega$ that addresses the $m_F=4\to m_{F'}=5$ transition, shown in the absence (dotted line) and in the presence (solid line) of a strong field that couples the $m_F=4\to m_{F'}=4$ transition.}
\label{fig:levels}
\end{figure}

A forward propagating light field which is aligned along the WCA is purely $\pi$-polarized at the position of the atoms and therefore only drives $\pi$-transitions between the two manifolds. Thus, the outermost Zeeman sub-states of the excited state $F'=5$ manifold are not addressed, see Fig.~\ref{fig:levels}(a). Still, the situation reduces to that of the two-level-atom above if the backward propagating light field couples to the same excited states, i.e., if it is also aligned along the WCA and thus $\pi$-polarized at the position of the atoms. The situation changes if we consider the case where the backward propagating light field is aligned along the SCA: In this case, the field also couples to the unsaturated  $|F=4, m_{F}=\pm 4\rangle \to |F'=5, m_{F'}=\pm 5\rangle$-transitions, thereby leading to absorption, see Fig.~\ref{fig:levels}(a). Its effect is, however, negligible for two reasons: First, assuming an initially flat distribution of the populations of the Zeeman sub-states of the $F=4$ ground state manifold and neglecting optical pumping, only $1/18$ of the population will be present in the $F=4, m_{F}= 4$ and the $F=4, m_{F}=- 4$ levels, respectively. Second, the strong forward propagating light field leads to an Autler-Townes splitting of these levels, see Fig.~\ref{fig:levels}(b). This further reduces the resonant absorption $A_0$ of the backward propagating field, see Fig.~\ref{fig:levels}(c). At large saturation, the latter scales as~\cite{Agarwal97}
\begin{figure}
\centerline{\includegraphics[width=0.95\columnwidth]{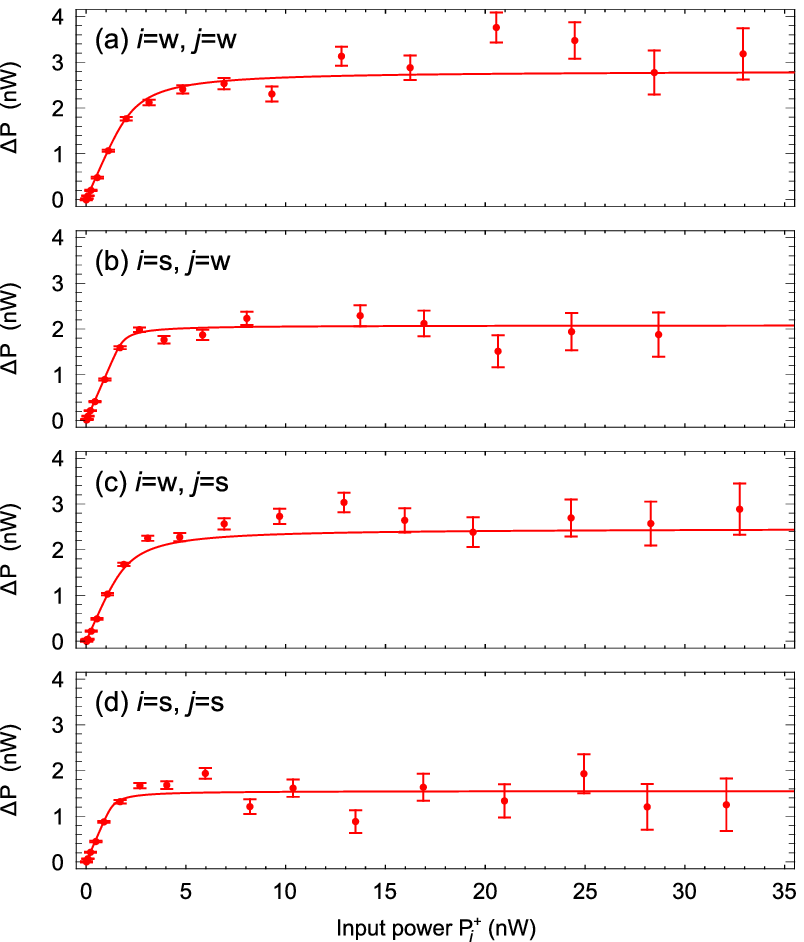}}
\caption{Difference $\Delta P$ between the transmitted powers with and without atoms trapped around the nanofiber as a function of the forward propagating power $P_i^+$. Each data point is averaged over 80 experimental runs.}
\label{fig:transmissionAll}
\end{figure}
\begin{equation}\label{eq:S1}
A_0\propto \left(1+\frac{4\Omega(z)^2}{9\Gamma^2}\right)^{-1},
\end{equation}
where $\Gamma$ is the excited state decay rate and $\Omega(z)$ is the Rabi frequency of the forward propagating field at position $z$. Using $\Omega(z)^2/\Gamma^2=\mathcal{P}_{i}^+(z) /2P_i^{\rm sat}=s_i(z)/2$, where $s_i(z)$ is the position-dependent saturation parameter, Eq.~\eqref{eq:S1} becomes $A_0\propto (1+2 s_i(z)/9)^{-1}$. The power-dependence is similar to that of the first term of Eq.~(1) of the main manuscript, and $A_0$ vanishes for large saturation parameters. Thus, while our model does not take this residual absorption into account, our conclusions in the strongly saturated regime obtained in the main text are valid for all polarizer--analyzer settings. Moreover, our model fits the data well even for small saturation.

\bigskip
\paragraph*{Calculation of the saturation power}
For a multi-level atom, the saturation intensity depends on the driven optical transitions~\cite{Steck10}. Furthermore, it is modified by Zeeman state-dependent light shifts induced by the trapping fields~\cite{Vetsch10}. Assuming an equal population of all Zeeman sub-states of the $F=4$ manifold and taking into account the effective mode area, we find $P^{\rm sat}_{\rm w} = 480\,{\rm pW}$  and $P^{\rm sat}_{\rm s} = 130\,{\rm pW}$, which enter the saturation parameter $s_i(z)$ with $i\in\{{\rm w},{\rm s}\}$ in the main manuscript. 

\section*{Transmission}
For completeness, in Fig.~\ref{fig:transmissionAll}, we show the transmission measurements for all four polarizer--analyzer settings. The fit results are used to determine the corresponding number of atoms, see Tab.~I in the manuscript.

\bibliography{Reflection}